\documentclass[aps,prb, twocolumn, superscriptaddress]{revtex4-2}
\usepackage{amsmath}
\usepackage[english]{babel}
\usepackage[pdftex]{graphicx}
\usepackage{color}
\usepackage{ulem}

\usepackage{hyperref}
\graphicspath{{images/}{./}}

\usepackage{amsmath}
\usepackage{amssymb}
\usepackage{mathtools}
\usepackage{braket}

\renewcommand{\vec}[1]{\ensuremath{\boldsymbol{#1}}}

\begin{document}

\title{Kekulé distortions in graphene on cadmium sulfide}

\author{Yonatan Betancur-Ocampo}
\email{ybetancur@fisica.unam.mx}
\affiliation{Instituto de F\'isica, Universidad Nacional Aut\'onoma de M\'exico, 04510, Ciudad de M\'exico, M\'exico}

\author{Santiago Galv\'an y Garc\'ia}
%\email{santiagogyg@icf.unam.mx}
\affiliation{Instituto de Ciencias F\'isicas, Universidad Nacional Aut\'onoma de M\'exico, 62210 Cuernavaca, M\'exico}

\author{Samuel Tehuacanero-Nu\~nez}
\affiliation{Instituto de F\'isica, Universidad Nacional Aut\'onoma de M\'exico, 04510, Ciudad de M\'exico, M\'exico}

\author{José Reyes-Gasga}
%\email{jreyes@fisica.unam.mx}
\affiliation{Instituto de F\'isica, Universidad Nacional Aut\'onoma de M\'exico, 04510, Ciudad de M\'exico, M\'exico}

\author{Thomas Stegmann}
%\email{stegmann@icf.unam.mx}
\affiliation{Instituto de Ciencias F\'isicas, Universidad Nacional Aut\'onoma de M\'exico, 62210 Cuernavaca, M\'exico}
%\phone{+52-777-3291-728}

\author{Francisco Sánchez-Ochoa}
%\email{fsanchez@fisica.unam.mx}
\affiliation{Instituto de F\'isica, Universidad Nacional Aut\'onoma de M\'exico, 04510, Ciudad de M\'exico, M\'exico}

\begin{abstract}
The deposition of a two-dimensional material on the surface of a three-dimensional crystal can generate superlattices with electronic properties modified through the proximity spin-orbit effect. In this study, we found that Kekulé graphene superlattices are obtained by placing graphene on Cd-terminated (001)-(1$\times$1) cadmium sulfide (CdS) surface. From an effective model of Kekulé superlattices, which is corroborated by Density Functional Theory (DFT) calculations, we identified that the puckered surfqace of CdS modifies the on-site energies (staggered potential) and C-C bonds, giving rise to two possible $\sqrt{3} \times \sqrt{3}$ hexagonal superlattices, which are known as Kekulé-O graphene and Kekulé-Y with a quadratic band crossing point. Both Kekulé superlattices present spin-orbit coupling due to the interaction of graphene with the CdS surface. To guide the experimental realization of Kekulé superlattices based on graphene/CdS heterostructures, we simulate electron diffraction patterns, as well as images from High-Resolution Transmission Electron Microscopy (HRTEM), and Scanning Transmission Electron Microscopy in the mode of High Angle Annular Dark Field (STEM-HAADF).
\end{abstract}

\maketitle

\section{Introduction}

The plethora of different electronic phases in two-dimensional materials has recently increased even further with the generation of superlattices in those materials \cite{Ponomarenko2013, Cao2018, Mao2020, Cao2018a, Bao2021, Garcia2022, sanchez2023, Cheianov2009, Gutierrez2016, Giovannetti2015, Jia2016, Venderbos2016, Skurativska2021, Wallbank2013}. Most prominently, the creation of moir\'e patterns served to discover unconventional superconductivity and Mott-like insulator states in twisted bilayer graphene \cite{Cao2018,Cao2018a}. This same technique was extended quickly for all two-dimensional materials such as twisted bilayer black phosphorus and transition metal dichalcogenides with the possibility to find out exotic and unusual physical properties \cite{Zhao2021,Fang2019,Weston2020,Wang2020a}. Most of the interest about this hot topic is the possibility to design novel nanodevices for spintronics, valleytronics, and twistronics applications. 

A subset of honeycomb superlattices is Kekulé graphene, which possesses a unit cell three times larger as in pristine graphene  \cite{Cheianov2009, Gamayun2018, Andrade2019, RuizTijerina2019, GonzalezArraga2018, Giovannetti2015, Gutierrez2016, Garcia2022, Bao2021, GonzalezArraga2018}. This allows us to couple the pseudo-spin valley degree of freedom with the one of the sublattice, very similar to Rashba or spin-orbit coupling interactions in spintronics \cite{Gamayun2018, Garcia2022, Santacruz2022, Andrade2022, Bercioux2015, Avsar2020, Pachoud2014, Wang2020, Vitale2018, Schaibley2016, Zeng2021, Alsharari2016, Kochan2017, Gmitra2017, Zhang2023}. Spin-orbit coupling has been addressed extensively in graphene due to the possibility to open a gap \cite{Konschuh2010}. However, this band gap is tiny for pristine graphene \cite{Kane2005}. From ab-initio studies the estimation of the gap is in the order of $\mu$eV \cite{Yao2007}. Other alternative ways are explored, such as depositing graphene on substrates or by considering adatoms, where the gap opening increases from around $ 100 \mu$eV to some few meV \cite{Gmitra2017, Weeks2011, Balakrishnan2013, Gmitra2013}. The strategies to get a Kekulé superlattice are multiple, among them and the most recent, is the intercalation of Li atoms in a wafer of bilayer graphene on SiC substrate \cite{Bao2021, Andersen2014, Watcharinyanon2012, Hsu2012, Yazdi2016, Forti2014, Halle2016, Hsu2013, Papagno2011, Pervan2015}. Atomically thin current pathways can be obtained by manipulating Ti atoms on graphene to create Kekulé textures, which is a technique that was named "Kekulé engineering" \cite{Garcia2024}. The wide variety of topological phases, which have been predicted and some tested experimentally, leads to propose other alternatives to create Kekulé superlattices, which include synthesized materials and artificial lattices \cite{Xie2019, Radha2020, Toyama2021, Shi2015, Freeney2020, Freeney2022}. Using scanning tunneling microscope (STM) to adsorb carbon monoxide molecules on Cu substrate made it possible to test the Kekulé-Y phase \cite{Gutierrez2016}. Photonic and acoustic waveguides are also platforms for the realization of Kekulé phases and helical edge states, where the manipulation of sites and coupling among resonators can be performed directly \cite{Mu2022, Freeney2022}. Other versions of Kekulé-Y exist without experimental test yet, where the electronic band structure present a couple of pseudo-relativistic dispersion relations, valley-dependent gap opening, and two quadratic bands that cross in a single point. This phase is known as quadratic band crossing point (QBCP) and emerges in hexagonal superlattices \cite{Giovannetti2015}, checkerboard \cite{Sun2009}, and Kagome lattices \cite{Du2017}, which may lead to the realization of unconventional superconductivity.

In this paper, we propose to deposite graphene on a CdS substrate in order to generate Kekulé superlattices. Specific orientations of the crystalline plane of CdS allow us to obtain $\sqrt{3} \times \sqrt{3}$ hexagonal superlattices, where the proximity spin-orbit effect of graphene on the CdS substrate gives rise to two phases of Kekulé graphene, which are named Kekulé-O (Kek-O) and Kekulé-Y (Kek-Y) with a quadratic band crossing point (QBCP). We study theoretically the electronic band structure of graphene on CdS by DFT and compare to an effective low-energy model, that shows excellent agreement. The Cd atoms induce a spin-orbit coupling in both phases of Kekulé graphene with a spin polarization oriented along the plane, which is independent of the wave vector. The effective model describes this polarization through a Rashba term that couples the three pseudospin degree of freedom: sublattice, valley, and spin. The Kekulé graphene phases, such as Kek-O and QBCP, which are based on the proximity spin-orbit effect with the CdS substrates, pave the way for the study of spin transistor and non-conventional superconductivity. To guide the experimental realization, we simulate the diffraction patterns of graphene on CdS obtained from Transmission Electron Microscopy (TEM).

\section{Atomic model \& DFT calculations}

\begin{figure*}
    \centering
    \includegraphics[scale=0.45]{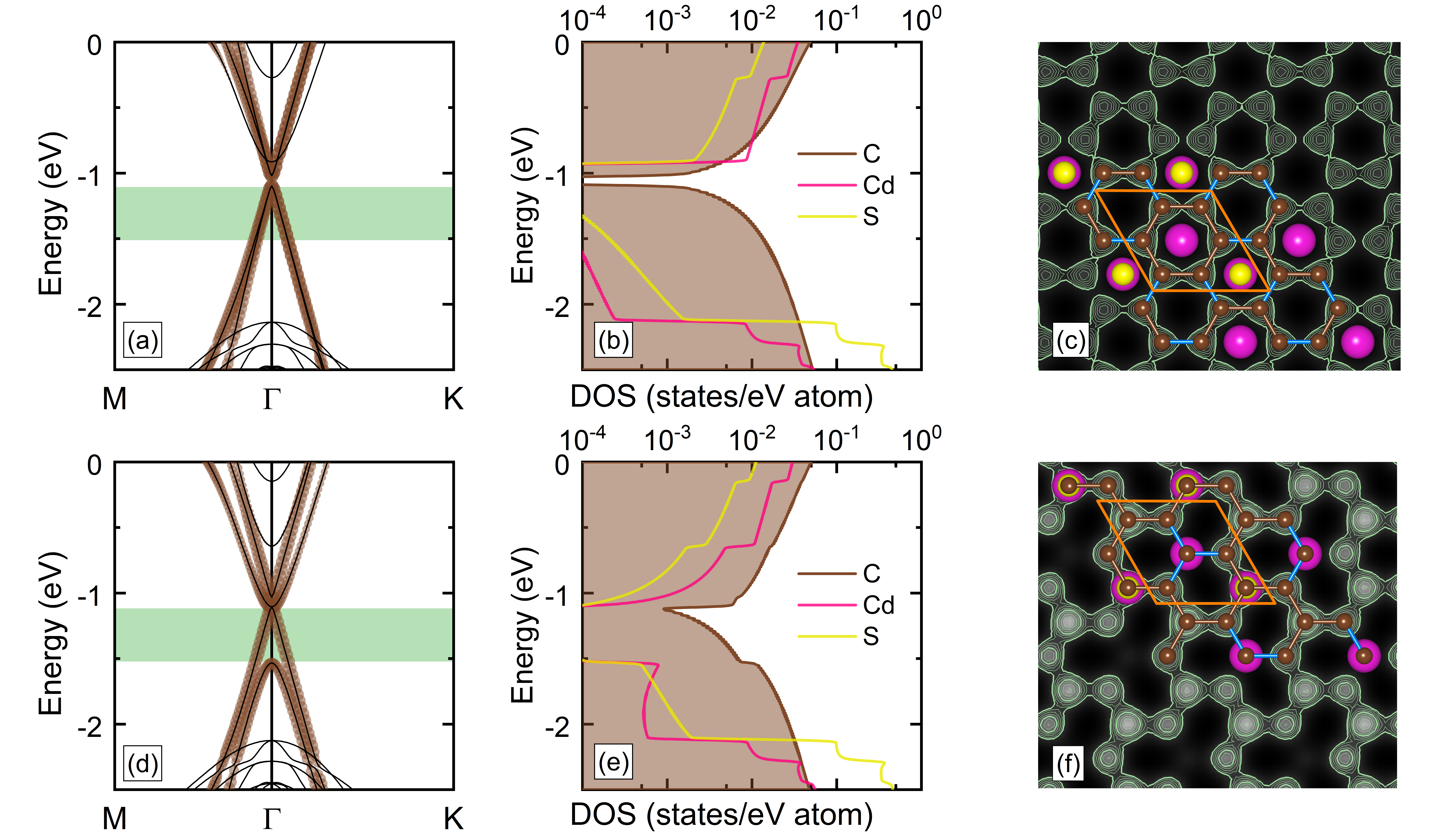}
    \caption{(a) Electronic band structure showing carbon contributions denoted by brown thick bands,  (b) Log-scale atom-resolved density of states (DOS) and (c) two dimensional charge densities calculated from local density of states (LDOS) for those states enclosed by the green rectangle in (a) for the Kek-O configuration. (d), (e) and (f) are the same but for QBCP configuration. The Fermi energy is set to zero eV. Pink, yellow and brown spheres represent Cd, S and C atoms. The orange lines in (c) and (f) denote the periodic graphene/CdS supercell. The bands gap for CdS systems are 1.21 eV and 1.47 eV for Kek-O and QBCP, respectively.}
    \label{atomicmodels}
\end{figure*}

Kekulé superlattices have been investigated recently due to the effect of folding of Dirac cones at the center of the first Brillouin zone, which have a strong resemblance with the effect of spin-orbit interactions on the electronic band structure of materials \cite{Cheianov2009,Gamayun2018,Garcia2022,Santacruz2022,Herrera2020}. The Kekulé superlattice emerges because the ratio of lattice constants of CdS and graphene is roughly $\sqrt{3}$, and therefore, the unit cell of the superlattice is three times the one of graphene. The opposite effect occurs on the reciprocal lattice, where the first Brillouin zone of the superlattice is reduced three times. This can be viewed as the folding of the original first Brillouin zone of graphene. This folding allows us to couple the pseudo-spin valleys. In this way, phenomena from spintronics is expected to be realized in Kekulé superlattices with the advantage that valley degree of freedom can be manipulated for valley-polarization currents \cite{Gamayun2018,Garcia2022}. 

The study of their electronic band structure can be performed from total energy DFT calculations using the SIESTA software \cite{soler} with norm-conserving pseudopotentials. The van der Waals density functional within the Klime$\check{s}$, Bowler, and Michaelides (KBM) parametrization for exchange and correlation effects were applied \cite{Klime}. Structural parameters and forces were well-converged using a 45$\times$45$\times$1 Monkhorst-Pack grid \cite{Monkhorst1976} and a 300 Ry meshcutoff. A Gaussian smearing of 100 K was used for the Fermi surface broadening. The optimization of atomic positions and lattice vectors were performed until residual forces were equal to or less than 10 meV/\AA; while the electronic relaxation was converged to 10$^{-5}$ eV. To simulate the graphene/CdS system, we employed the supercell method with a vacuum space of 20~\AA~to avoid replica interactions. Slab-dipole corrections were included to compensate the dipole generated by the slab \cite{Bengtsson}. The visualization of atomic models and charge densities was performed using the VESTA program \cite{Momma2011}. The S-terminated (001) CdS surface, see Fig. S1 of Supplementary Information (SI) file, was passivated with fractionally charged ‘‘pseudohydrogen’’ atoms. More details of fractional H charges passivating the S-terminated (00$\bar{1}$) surface and the electronic structure are presented in Fig. S2 of SI file. Aditionally, the number of CdS layers as well as the corresponding changes in electronic structure of graphene with more CdS layers were systematically studied and the results are shown in Fig. S3-S4 (SI). Hence, six CdS layers plus H$^{0.5+}$ atoms (passivating the opposite S-terminated surface), are enough to simulate the Cd-terminated (001) CdS surface.  

Optimized lattice constants for the bulk CdS unit cell in the wurtzite phase are $a_{\mathrm{CdS}}$= 4.094~\AA~and $c_{\mathrm{CdS}}$=6.655~\AA~keeping the P6$_\text{3}$mc space group, while for $\sqrt{3} \times \sqrt{3}$ graphene monolayer the lattice constant is $a_\mathrm{G}$= 4.261 \AA. Thus, the lattice mismatch ($a_{\mathrm{G}}$-$a_{\mathrm{CdS}}$)/$a_{\mathrm{CdS}}$ is 4.08$\%$ within the exchange-correlation approximation. Once the pristine graphene monolayer and the CdS bulk structures were obtained, we built a Cd-terminated (001) CdS surface unit cell (1$\times$1) to adsorb the $\sqrt{3} \times \sqrt{3}$ graphene monolayer. The Cd-terminated (001) CdS surface is ideal to adsorb graphene since C atoms preferentially adsorb on top of Cd atoms \citenum{Manade2015}, in addition to the mismatch cell above mentioned. A relative energy difference between the two Kekulé phases, Kek-O and QBCP, of 22.3 meV per unit cell of graphene is obtained; being the QBCP phase the most stable structure. The average C-Cd distance is 3.06 \AA~and 2.84 \AA~for Kek-O and QBCP, respectively, showing a strong Cd-C interaction for the QBCP configuration. Indeed, we observe a small deviation of 0.4$\%$ between all the C-C distances in the QBCP system meanwhile there are no changes in C-C distances for the Kek-O system. These results show that hopping parameters between C(2p$_z$) orbitals could be modified by on-site energies of C atoms by Cd atoms.

\section{Energy bands and Density of States}

\begin{figure*}
    \centering
    \includegraphics[scale=0.4]{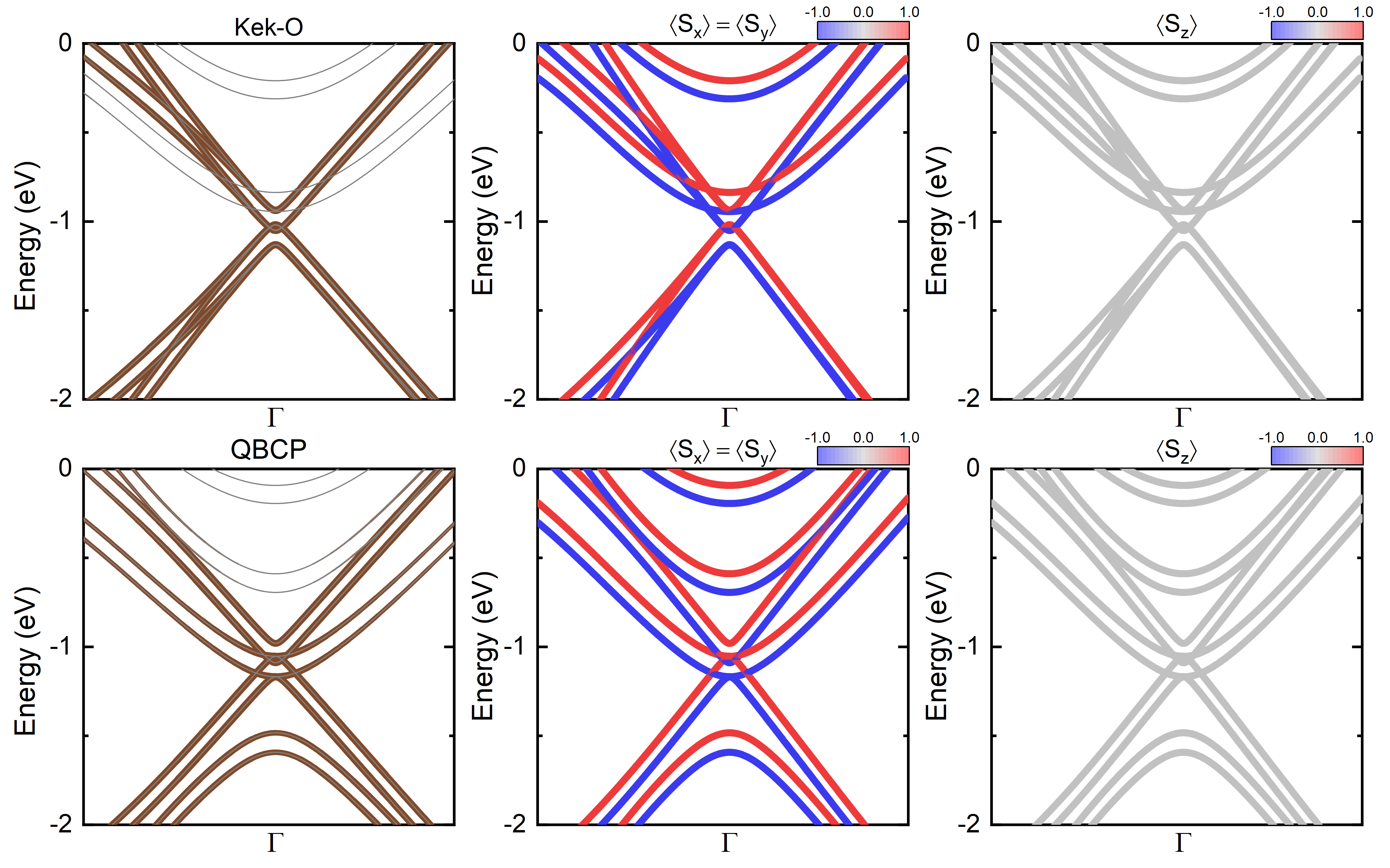}
    \caption{Top panel: Thick bands show the carbon contributions with brown circles and spin projections for Kek-O graphene system. Bottom panel: The same as top panel but for QBCP configuration. The Fermi energy is set to zero eV.}
    \label{soc}
\end{figure*}

Figures \ref{atomicmodels}(a) and (b) show the band structure and partial DOS of the electron-doped Kek-O system where a small bandgap of $\sim80$ meV is observed below -1 eV. The Dirac point in this case is pined to the first unoccupied band from CdS. Two degenerated bands from graphene are enclosed by the band gap of CdS from -2.12 eV to -0.91 eV. Figure \ref{atomicmodels}(c) presents the local density of states (LDOS) calculated for those electronic states enclosed by the green rectangle in Fig. \ref{atomicmodels}(a), with solely graphene contributions; but indeed revealing isolated benzene rings. This is a clear observation of a Kek-O configuration in graphene induced by a Cd-terminated (001) CdS surface. Regarding the electronic structure of the QBCP configuration, see Fig. \ref{atomicmodels}(d) to (f),  the band degeneracy is lifted with two couple of bands identical to the ones of QBCP, which are pseudo-relativistic dispersion relations with two different relative energy gaps. A bandgap corresponds to $\sim80$ meV and the other is $\sim0.43$ eV. Notice that, two bands touch at -1.1 eV giving a metallic behavior to graphene. Such that, the full system is composed of metallic properties by graphene enclosed by the semiconductor behavior of CdS from -2.11 eV to -0.64 eV. in In Fig. \ref{atomicmodels}(f), the calculated LDOS using the enclosed states by the green rectangle shows a clear charge distribution with a Kek-Y pattern in real space in 2D. To end this section, we show in Fig. \ref{soc} the band structure of two configurations with carbon and spin contributions taking into account the spin-orbit coupling effect \cite{Cuadrado}. Spin projection indicates that there is an intrinsic polarization in the direction $\theta = \pi/4$ in the $xy$ plane, as shown in Fig. \ref{soc}. Such a magnetization is due to the puckered surface of CdS substrates. In forthcoming section, we investigated this spin-polarization from an effective model which considers the coupling of the electron spin with the valley and sublattice pseudospin.

\section{Effective model for Kekulé superlattices}

\begin{figure*}[t!!]
    \centering
    \begin{tabular}{c c}
           \includegraphics[scale=0.27]{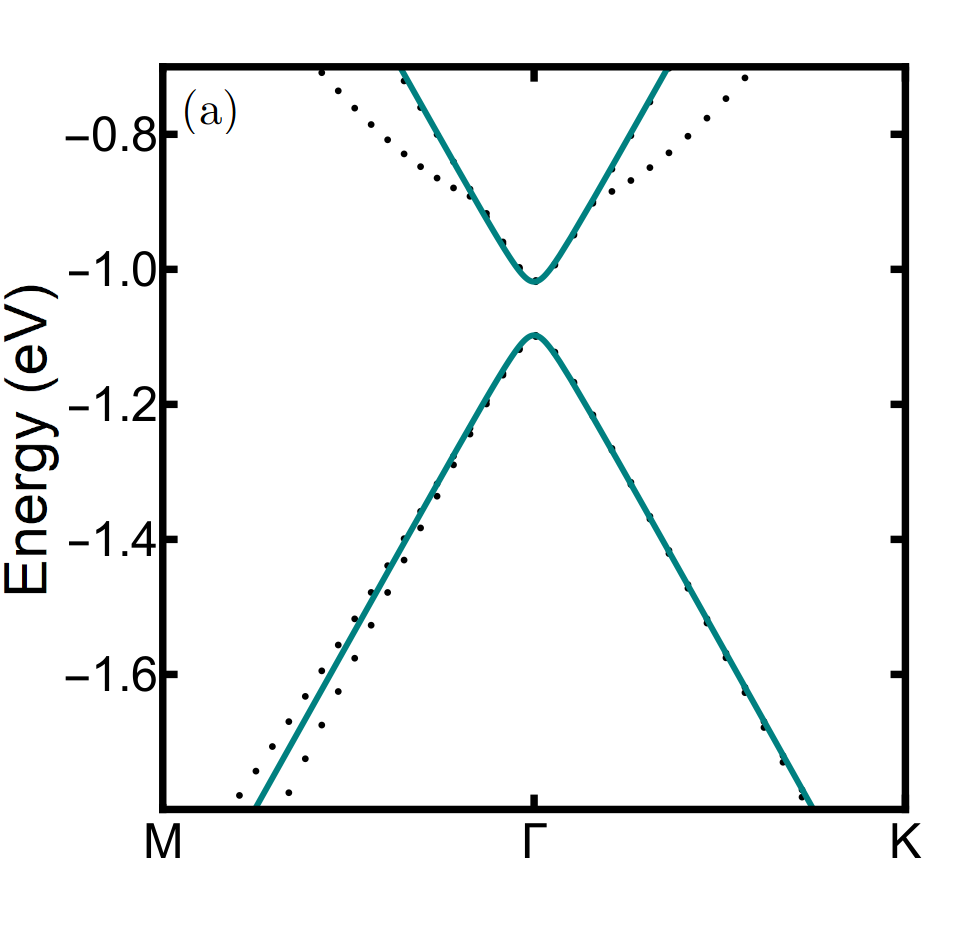} &
          \includegraphics[scale=0.27]{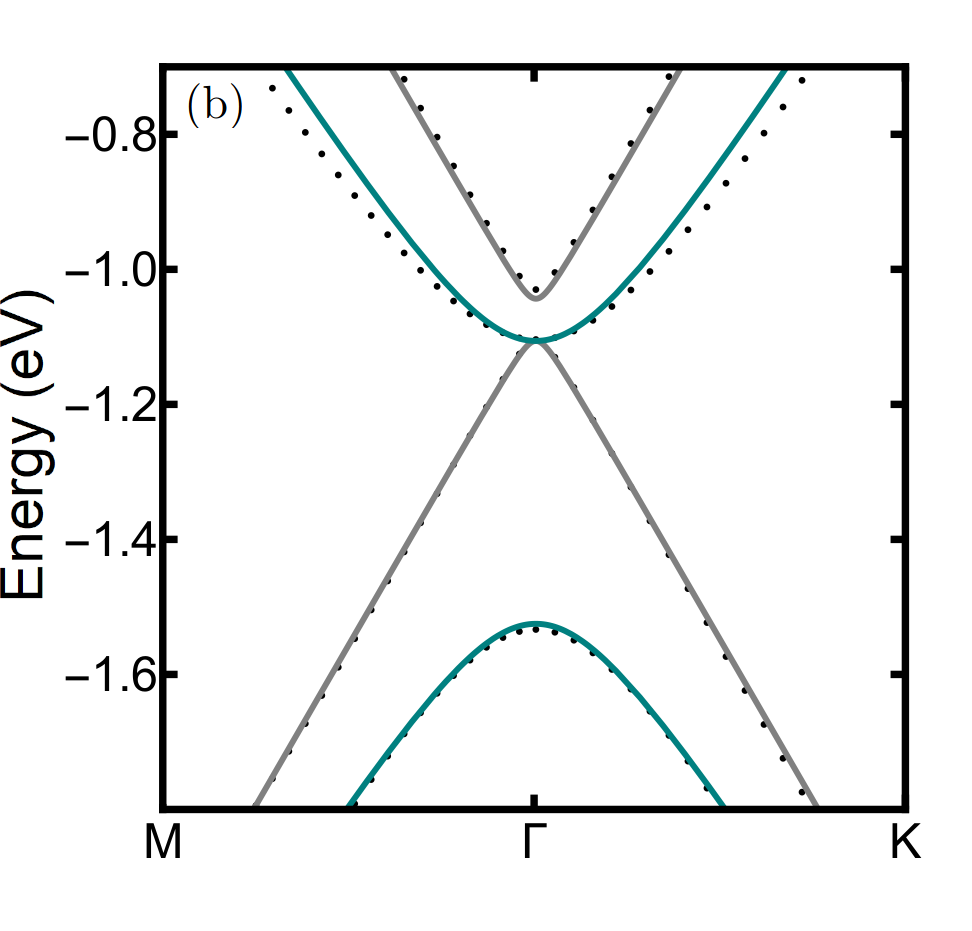}\\
          \includegraphics[scale=0.281]{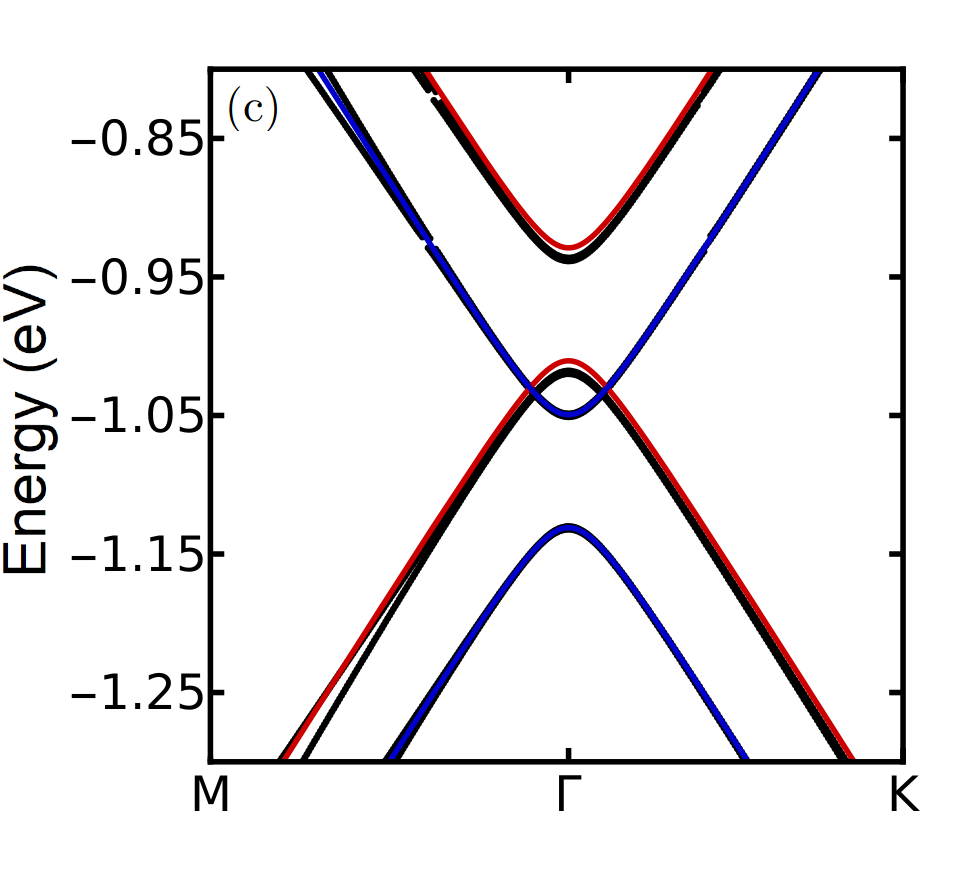} &
          \includegraphics[scale=0.27]{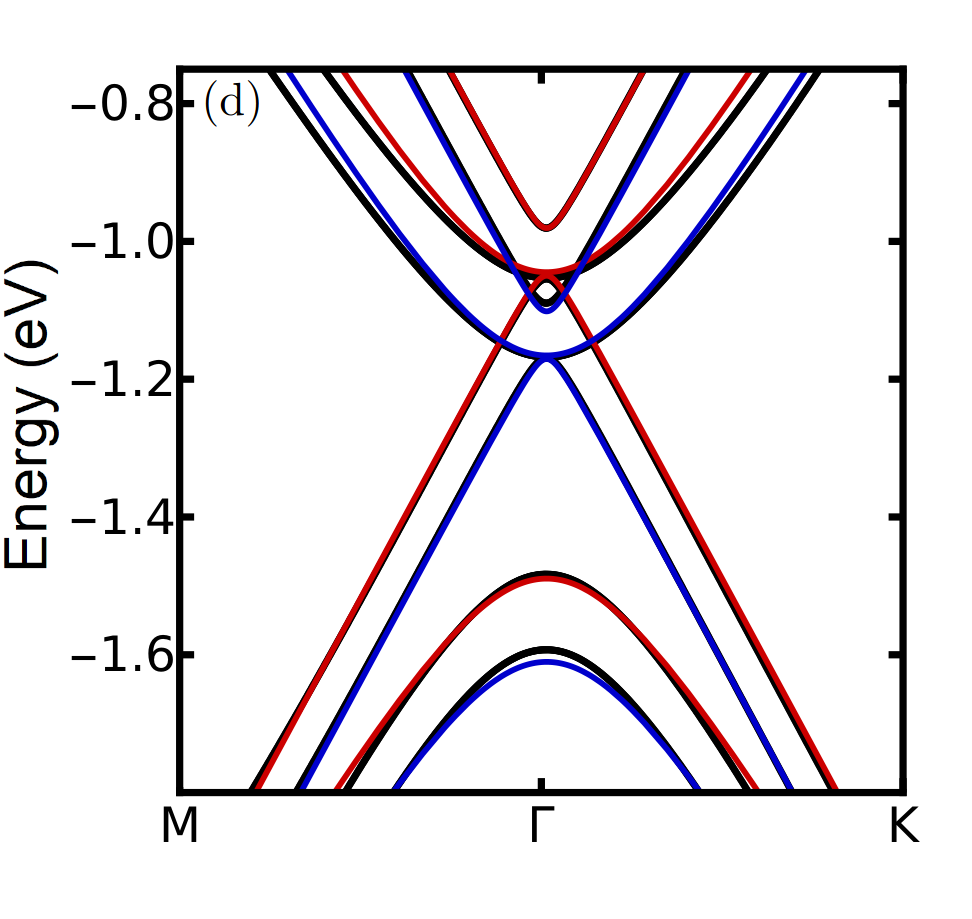}
    \end{tabular}
    \caption{Comparison between DFT calculations (dotted black points) and effective model for the electronic band structure of Kekulé graphene for (a) Kekulé-O graphene superlattice and (b) quadratic band crossing point. The best fit of the dispersion relation by using Eq. \eqref{drKEKO} is $\mu = 39.7$ meV and $v_\sigma = 7.91 \times 10^5\, \textrm{ms}^{-1}$. From Eq. \eqref{drQBCP}, the best fit is $\mu_{-}=22$ meV, $v_{-}=7.57 \times 10^5\,\textrm{ms}^{-1}$ for the valley $K^{-}$, and $\mu_{+}=209$ meV, $v_{+}=7.15 \times 10^5\,\textrm{ms}^{-1}$ for the valley $K^+$. Spin-orbit coupling effects for Kekulé-O graphene in (c) and quadratic band crossing point in (d). In the Kek-O phase, the best fit parameters are $v_\sigma = 8.0\times 10^5\,\textrm{ms}^{-1}$, $\mu = 40.78$ meV, and quadratic band crossing point are $\mu_- = 11.97$ meV, $v_- = 7.84 \times 10^5\,\textrm{ms}^{-1}$, $\mu_+ = 280$ meV, and $v_+ = 7.13 \times 10^5\,\textrm{ms}^{-1}$. The spin-orbit strength coupling is $\lambda = 60.34$ meV.}
    \label{FEBs}
\end{figure*}

 The electronic band structure of Kekulé superlattices can also be studied by effective models based on the tight binding approach to nearest neighbors \cite{Gamayun2018,Andrade2019,Garcia2022}. Two Kekulé graphene phases of interest in the present work are depicted by two effective Dirac-like Hamiltonians in the continuum approximation \cite{Garcia2022}. The first phase for Kek-O, the effective Hamiltonian is
\begin{eqnarray}
    H_\textrm{Kek-O} & = & \left(\begin{matrix}
    0 & v_\sigma p_- & \mu & 0\\
     v_\sigma p_+ & 0 & 0 & -\mu\\
     \mu & 0 & 0 & v_\sigma p_-\\
     0 & -\mu & v_\sigma p_+ & 0
    \end{matrix}\right),
    \label{HKEKO}
\end{eqnarray}

\noindent where the linear momentum is $\vec{p} = (p_x,p_y)$ and we define $p_\pm = p_x \pm ip_y$. The Fermi velocity is $v_\sigma$. The band structure is obtained by diagonalizing the Hamiltonian $H_\textrm{Kek-O}$ in Eq. \eqref{HKEKO},

\begin{equation}\label{drKEKO}
    E^\textrm{Kek-O}_s=s\sqrt{v_\sigma^2p^2+\mu^2}
\end{equation}

\noindent with a twofold valley-generated energy bands. In the second phase, which corresponds to Kek-Y graphene with a QBCP, the effective Hamiltonian is

\begin{eqnarray}
    H_\textrm{QBCP} & = & \left(\begin{matrix}
    0 & v_\sigma p_- & v_\tau p_- & 0\\
     v_\sigma p_+ & 2\rho & 2\delta & v_\tau p_-\\
     v_\tau p_+ & 2\delta & 2\rho & v_\sigma p_-\\
     0 & v_\tau p_+ & v_\sigma p_+ & 0
    \end{matrix}\right),
    \label{HQBCP}
\end{eqnarray}

\noindent where $v_\tau$ quantifies the Kekulé distortion. The quantities $\rho$ and $\delta$ are effective parameters of the model, which are related with the valley-dependent gap opening. The dispersion relation is given by

\begin{eqnarray}\label{drQBCP}
    E^\nu_s & = & \mu_\nu + s\sqrt{v_\nu^2p^2 + \mu_\nu^2}.
\end{eqnarray}

\noindent We define the quantities $\mu_\nu =\rho + \nu\delta$ as the rest energy, $v_\nu = v_\sigma+\nu v_\tau$ is the modified Fermi velocity, $s = \pm 1$ is the band index and $p = \sqrt{p_x^2 + p_y^2}$ is the magnitude of the linear momentum, and $\nu = \pm 1$ is the valley index. The energies $E_s^-$ and $E_s^+$ are the folded bands of the mapped valleys $K^-$ and $K^+$, respectively. In this QBCP, electrons behave as massive Dirac fermions with valley-dependent limit velocities $v_\sigma + \nu v_\tau$, relative band shifts and rest energies $\rho + \nu\delta$. The QBCP occurs when $p = 0$ in the bands $E^-_-$ and $E^+_-$.

The explanation for the origin of the two phases described by the effective Hamiltonians \eqref{HKEKO} and \eqref{HQBCP} is due to the proximity effect \cite{Garcia2022}. We noted that the superlattice configuration, as shown in Fig. \ref{atomicmodels}(f) is obtained by depositing graphene on a puckered surface of the CdS substrate. By this proximity effect, both S and Cd atoms affect straightforwardly the on-site energy of the C atoms and nearest neighbor hopping parameters, which causes a periodically distortion in graphene and gives rise to the Kekulé superlattice. Due to that the S atoms are closer to the C atoms, the on-site energy and nearest-neighbor hopping parameters of graphene are modified larger than the distortion caused by Cd atoms. The effective parameters $\mu_\nu$ and $v_\nu$ of the Hamiltonian in Eq. \eqref{HQBCP} quantify the grade of distortion by CdS substrate for the QBCP phase. The values of these parameters can be determined by fitting the energy bands, which are obtained from DFT calculations, as shown in Fig. \ref{FEBs}(a). We found that there is a good agreement of the dispersion relations in Eq. \eqref{drQBCP} with the energy bands of DFT calculations, where electrons in valley $K^-$ have the effective rest energy $\mu_- = 2.21$ meV and $v_- = 7.57 \times 10^5$ ms$^{-1}$, and for the valley $K^+$, $\mu_+ = 20.9$ meV and $v_+ = 7.15 \times 10^5$ ms$^{-1}$. However, it is important to note that the fit of the dispersion relations in Eqs. \eqref{drQBCP} and energy band is not perfect because there is an electron-hole asymmetry. Such an asymmetry appears by the electron-electron interactions, which is taken into account in the DFT calculation, while the effective Hamiltonian in Eq. \eqref{HQBCP} depicts the dynamics of a single particle. 

The second superlattice corresponds to Kek-O phase and emerges if the Cd atoms are located in alternating benzene rings of the graphene monolayer [see Fig. \ref{atomicmodels}(c)]. Bond distortions in benzene rings depend on the equilibrium distance of graphene on CdS substrate. Kek-O texture is obtained by depositing graphene in the crystalline orientation, as shown in Fig. \ref{atomicmodels}(c). By fitting the dispersion relation in Eq. \eqref{drKEKO} with the energy bands from the DFT calculations, we can see that there is a qualitative agreement within low-energy regime [see Fig. \ref{FEBs}(b)]. The Cd and S atoms are located at the center of the benzene rings and open a gap in graphene, which is charateristic of Kek-O superlattices \cite{Gamayun2018,Bao2021}. As in the QBCP phase, an electron-hole asymmetry is also observed in the Kek-O phase with DFT calculations, and it is due to electron-electron interactions. This asymmetry is more evident for energies far away from the gap.

 To include the proximity spin-orbit coupling of graphene induced by CdS substrates, the addition of the Rashba term in Hamiltonians of Eqs. \eqref{HKEKO} and \eqref{HQBCP}

\begin{eqnarray}
    H_\textrm{SOC} & = & \frac{\lambda}{\sqrt{2}}(s_x+s_y)\otimes\tau_0\otimes\sigma_0
\end{eqnarray}

\noindent gives rise to spin-dependent energy bands of graphene/CdS heterostructure, where $\lambda$ is the strength of the spin-orbit coupling, $s_x$ and $s_y$ are the $x$ and $y$ components of Pauli matrix for the instrinsic spin electron, $\tau_0$ and $\sigma_0$ are the identity matrices for the pseudospin valley and sublattice, respectively. The energy bands in Eqs. \eqref{drKEKO} and \eqref{drQBCP} are modified by the spin-orbit coupling as

\begin{equation}
    E'_\textrm{Gr/CdS} = E_\textrm{KEKO/QBCP} \pm \lambda,
\end{equation}

\noindent where the sign $\pm$ creates a relative separation of $2\lambda$ for each one of the four bands in Eqs. \eqref{drKEKO} and \eqref{drQBCP} with spin-up (blue) and spin-down (red), as shown in Fig. \ref{FEBs}(c) and (d). 
The separation of the energy bands lifts their degeneracy and closes the band gap. The effective model coincides with the in-plane magnetization observed from DFT calculations in Fig. \ref{soc}, which takes into account spin-orbit coupling induced by CdS substrates. The spin polarization is oriented along the $\pi/4$ direction in the $xy$ plane for both phases KEK-O and QBCP. The fitted value of spin-orbit coupling strength is $\lambda \approx 60.34$ meV. 

\section{Diffraction patterns and Microscopy}

\begin{figure*}[t!!]
    \centering
    \includegraphics[scale=0.4]{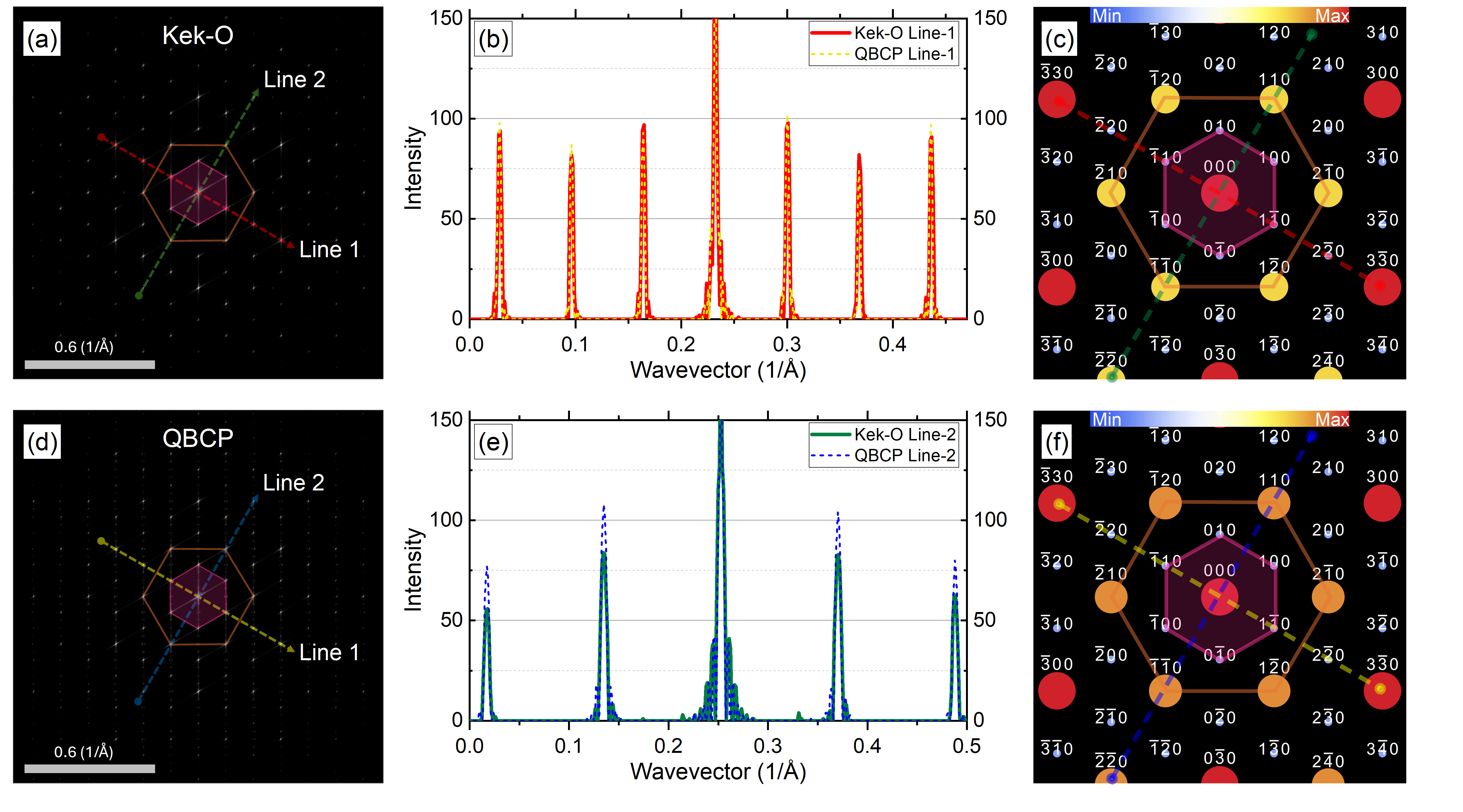}
    \caption{Simulated electron diffraction pattern in the [001] direction of the Graphene/CdS arrangements shown in Figure \ref{atomicmodels} with (a) Kek-O and (d) QBCP configuration. (b) and (e) are the corresponding intensity profiles. (c) and (f) show the simulated diffraction patterns normalized with the intensity of the $\Gamma$-point with Miller indices at 000.}
    \label{hrtem-haadf}
\end{figure*}

As a guide for experimentalists, here we present the theoretical results for two microscopy techniques based on 2D diffraction patterns, High resolution transmission electron microscopy (HRTEM) and High-angle annular dark-field (HAADF), which can help to identify the two Kekulé phases as well as the relative orientation between graphene and CdS and its corresponding periodicity.

The simulation of the electron diffraction pattern of graphene on CdS were employed using the SIMULATEM multilayer software \cite{GomezRodriguez2010}. HRTEM images were also simulated with this software, while HAADF images were carried out using the software STEM-CELL \cite{grillo2013}. These images are shown in the Fig. S5 of SI. The simulations were carried out for an HRTEM microscope with an accelerating voltage of 200 kV. The 200 kV HRTEM microscope was considered to have a spherical aberration coefficient C\textsubscript{s} of 0.5 mm, which implies a point-to-point resolution of 0.119 nm and Scherzer defocus of -15.81 nm. The results are shown in Figs. \ref{hrtem-haadf}(a) to (d), respectively, for the Kek-O and QBCP configurations. The graphene-CdS superlattice has the same periodicity of the CdS surface unit cell ($1\times1$). This is shown by the pink hexagonal cell delimited by the nearest reciprocal lattice points to the origin ($\Gamma$ point). On the other hand, the next-nearest reciprocal lattice points match with $\sqrt{3}\times\sqrt{3}$ graphene. Note that, a relative orientation of 30$^{\circ}$ is observed between two primitive cells of CdS ($1\times1$) and graphene ($\sqrt{3}\times\sqrt{3}$). Since the intensity of bright spots of the simulated electron diffraction pattern shown in Figs. \ref{hrtem-haadf}(a) and (d) is very weak, it is hard to identify differences between two configurations, at first insight. We calculate two orthogonal intensity profiles for each configuration, named Line 1 and Line 2. Fig. \ref{hrtem-haadf}(b) shows a comparison between intensity profiles results for Kek-O and QBCP along the Line 1. As shown, the seven peaks have in average the same height. Regarding the intensity along Line 2, for both configurations in Fig. \ref{hrtem-haadf}(e), a clear difference is observed between first and second peaks around the central peak at 0.25 1/\AA, which is the $\Gamma$-point. 

In order to get a clearer comparison between two electron diffraction patterns for Kek-O and QBCP configurations, we have calculated the structure factor, $F_{hkl}$, as follows:

\begin{equation}
    F_{hkl} = \sum_\textrm{$m$ atoms} f_m e^{i 2\pi \left( hx_m + hy_m + lz_m\right) },
\end{equation}

\noindent being $f_m$ the atomic form factor. Here $f_m$ is constant for any $G$ vector and equal to the charge number $Z$. $hkl$ are the integer Miller indices, and $x_m$, $y_m$ and $z_m$ are the fractional coordinates of the atomic basis. Additionally, we can index the reciprocal lattice points. The calculated diffraction patterns are present in Figs. \ref{hrtem-haadf}(c) and (f) for Kek-O and QBCP systems, respectively. A fair agreement is obtained with those patterns calculated with SIMULATEM, due to the match among the red, blue, green and yellow lines as well as the relative orientation between the CdS surface and graphene. The intensity of seven peaks, ($\bar{3}30$, $\bar{2}20$, $\bar{1}10$, $000$, $1\bar{1}0$, $2\bar{2}0$ and $3\bar{3}0$), along the Line 1 for Kek-O (dashed red line) and QBCP (dashed yellow line) is exactly the same. Conversely, the intensity of four peaks, ($\bar{2}\bar{2}0$, $\bar{1}\bar{1}0$, $110$ and $220$), along the Line 2 for Kek-O (dashed green line) and QBCP (dashed blue line) structures are different.

\section{Conclusions}
The electronic properties of graphene on CdS substrates can lead to the formation of Kekulé superlattices depending on the specific orientation of the crystalline structure of CdS. We identify two possibilities for Kekulé graphene superlattices on CdS substrates, which are Kekulé-O graphene (Kek-O) and quadratic band crossing point (QBCP). By using two methodologies one from DFT calculations and the other from an effective model of Kekulé superlattices, we study the proximity effect of graphene/CdS heterostructures by showing the contribution of pseudospin sublattice, valley, and spin-orbit coupling in the electronic band structure. The local density of states allows us to observe the charge distribution that characterizes Kek-O texture and QBCP. The effective model of Kekulé superlattices has an excellent agreement with DFT calculations and depicts the electron dynamics in graphene/CdS heterostructures. The QBCP is the most stable phase and possesses electrons that behave as valley-dependent massive Dirac fermions. Both phases, Kek-O and QBCP, present a band splitting due to the spin-orbit coupling. From DFT calculations, spin-projection on the band structure reveals that the spin direction is located along the direction $\theta = \pi/4$ in the $xy$ plane. Therefore, the present electronic properties study indicated that depositing graphene on CdS substrates may be an ideal platform for designing spintronics devices and experimental realization of unconventional superconductivity. Such a proposal is supported by simulating the systems Kekulé-O and QBCP from HRTEM and STEM-HAADF microscopy images of depositing graphene on CdS substrates in the crystalline orientations $[001]$ and $[100]$. 

\section{Acknowledgments}
We gratefully acknowledge financial support from UNAM-PAPIIT under project numbers IA106223, IA105623, IN103922 and CONAHCYT grant CF-2023-I-336. Numerical calculations were performed in Holiday cluster at Physics Institute (UNAM).
%
%\bibliography{Kek_CdS}
\end{document}